\documentclass[preprint,authoryear,12pt]{elsarticle}

\usepackage{latexsym}
\usepackage{graphicx} 
\usepackage{amsfonts}

\usepackage{subfigure}
\usepackage{float}

\usepackage{amsmath} %%\boldsymbol{}  %% \mathbf{ }
\usepackage{amssymb}
\usepackage{stmaryrd}
\usepackage{natbib}
\usepackage{enumerate}
\usepackage{amsmath}

\newcommand{\dfn}{\triangleq}
\usepackage{amsfonts}

\usepackage[usenames]{color}

\usepackage{amssymb}
\journal{Statistics $\&$ Probability Letters}

\begin{document}

\begin{frontmatter}

%% Title, authors and addresses

%% use the tnoteref command within \title for footnotes;
%% use the tnotetext command for the associated footnote;
%% use the fnref command within \author or \address for footnotes;
%% use the fntext command for the associated footnote;
%% use the corref command within \author for corresponding author footnotes;
%% use the cortext command for the associated footnote;
%% use the ead command for the email address,
%% and the form \ead[url] for the home page:
%%
%% \title{Title\tnoteref{label1}}
%% \tnotetext[label1]{}
%% \author{Name\corref{cor1}\fnref{label2}}
%% \ead{email address}
%% \ead[url]{home page}
%% \fntext[label2]{}
%% \cortext[cor1]{}
%% \address{Address\fnref{label3}}
%% \fntext[label3]{}

\title{A multi-point Metropolis scheme with generic weight functions}

%% use optional labels to link authors explicitly to addresses:
%% \author[label1,label2]{<author name>}
%% \address[label1]{<address>}
%% \address[label2]{<address>}

\author{Luca Martino, Victor Pascual Del Olmo, Jesse Read}

\address{ Department of Signal Theory and Communications, \\
Universidad Carlos III de Madrid. \\
Avenida de la Universidad 30, 28911, Leganes, Madrid, Spain. \\
{\tt luca@tsc.uc3m.es}, {\tt  vpolmo@tsc.uc3m.es}, {\tt jesse@tsc.uc3m.es} } %E-mail: {\tt luca@tsc.uc3m.es}}}

\begin{abstract}
The multi-point Metropolis algorithm %of \citep{Qin01} 
is an advanced MCMC technique based on drawing several correlated samples at each step and 
choosing one of them according to some normalized weights. We propose a variation of this technique where the weight functions are not specified, i.e., the analytic form can be chosen arbitrarily. This has the advantage of greater flexibility in the design of high-performance MCMC samplers. We prove that our method fulfills the balance condition, and provide a numerical simulation. We also give new insight into the functionality of different MCMC algorithms, and the connections between them.
%The new algorithm fulfills the detailed 
\end{abstract}
%% keywords here, in the form: keyword \sep keyword

%% MSC codes here, in the form: \MSC code \sep code
%% or \MSC[2008] code \sep code (2000 is the default)

\end{frontmatter}

% \linenumbers

%% main text

\section{Introduction}

Monte Carlo statistical methods are powerful tools for numerical inference and stochastic optimization (see \cite{Robert04}, for instance).
Markov chain Monte Carlo (MCMC) methods are classical Monte Carlo techniques that generate samples from a target probability density function (pdf) by drawing from a simpler proposed pdf, usually to approximate an otherwise-incalculable (analytically) integral \citep{Liu04b,Liang10}. MCMC algorithms produce a Markov chain with a stationary distribution that coincides with the target pdf.

The Metropolis-Hastings (MH) algorithm \citep{Metropolis53,Hastings70} is the most famous MCMC technique. It can be applied to almost any target distribution. In practice, however, finding a ``good'' proposal pdf can be difficult. In some applications, the Markov chain generated by the MH algorithm can remain trapped almost indefinitely in a local mode meaning that, in practice, convergence may not be reached. % \marginpar{e incluso infinito, verdad?}.

The {\it Multiple-Try Metropolis} (MTM) method of \cite{Liu00} is an extension of the MH algorithm in which the next state of the chain is selected among a \emph{set} of independent and identically distributed (i.i.d.) samples. This enables the MCMC sampler to make large step-size jumps without a lowering the acceptance rate; and thus MTM is can explore a larger portion of the sample space in fewer iterations. 

%%%% CAMBIO AQUÍ %%%%
An interesting special case of the MTM, well-known in molecular simulation field, is the {\it orientational bias Monte Carlo}, as described in Chapter 13 of \cite{Frenkel96} and Chapter 5 of \cite{Liu04b}, where i.i.d.\ candidates are drawn from a {\it symmetric} proposal pdf, and one of these is chosen according to some weights directly proportional to the target pdf. 
 %(evaluated in the generated samples). 
Here, however, the analytic form of the weight functions is fixed and unalterable.

\cite{Casarin2011} introduced a MTM scheme using different proposal pdfs. In this case the samples produced are independent but not identically distributed. In \cite{Qin01}, another generalization of the MTM (called the {\it multi-point Metropolis} method) is proposed using correlated candidates at each step. Clearly, the proposal pdfs are also different in this case. 

Moreover, in \cite{Pandolfi10} an extension of the classical MTM technique is introduced where the analytic form of the weights is not specified. In \cite{Pandolfi10}, the same proposal pdf is used to draw samples, so that the candidates generated each step of the algorithm are i.i.d.  %%% TENGO QUE PONER OTRO PUNTO ???
Further interesting and related considerations about the use of auxiliary variables for building acceptance probabilities within a MH approach can be found in \cite{Storvik11}.%\marginpar{las probabilidades son flexibles, or el metodo de construirlas?} 
 % (other related methods can be found in \citep{Frenkel, Doucet}).

%The multi-point scheme of \citep{Qin01} can seem similar to the MH algorithm with {\it delayed rejection} \citep{Green01,Mira01,Tierney99}. So much as some authors \citep{Storvik11} have considered the MTM a special case of the MH with delayed rejection but actually the two approaches are different. Indeed, in the delayed rejection scheme a MH-type test is performed to each candidate iteratively, while in the multi-point scheme all the samples are first stored and just one test is performed after the choise of a ``good'' sample.   

%but it is false...
%Some authors consider the MTM a special case of the Tierney and Mira .... 
%%%% CAMBIO AQUÍ %%%%

%In this work, we take from the two previous approaches to create a novel algorithm. Our algorithm selects a new state of the chain among correlated samples using generic 
In this paper, we draw from the two approaches \citep{Qin01} and \citep{Pandolfi10} to create a novel algorithm that selects a new state of the chain among correlated samples using generic weight functions, i.e., the analytic form of the weights can be chosen arbitrarily. Furthermore, we formulate the algorithm and the acceptance rule in order to fulfill the detailed balance condition.  

%%%%%%%%%%%%%%%%%%%%%%%%%%

Our method allows more flexibility in the design of efficient MCMC samplers with a larger coverage and faster exploration of the sample space. In fact, we can choose any bounded and positive weight functions to either improve performance or reduce computational complexity, independently of the chosen proposal pdf. % without checking any further conditions.... 
Moreover, since in our approach the proposal pdfs are different, adaptive or interacting techniques can be applied, such as those introduced by \cite{Andrieu06,Casarin2011}. 
An important advantage of our procedure is that, since in our procedure a new candidate is drawn from a conditional pdf which depends on the samples generated earlier during the same time step, it constructs an improved proposal by automatically building on the information obtained from the generated samples.

% according to the flexibility in the choice of the proposal and weight functions.    

%%%%%%%%%%%%%%%%%%%%%%%%%%

The rest of the paper is organized as follows. In Section \ref{SectMPM} we recall the standard multi-point Metropolis algorithm. In Section \ref{OurSect} we introduce our novel scheme with generic weight functions and correlated samples. Section \ref{SectProof} provides a rigorous proof that the novel scheme satisfies the detailed balance condition. 
% HE QUITADO ESTA FRASE In Section \ref{SpecCase} we show that the multi-point Metropolis method is a specific case of the novel scheme.
A numerical simulation is provided in Section \ref{ToyEx} and finally, in Section \ref{sectConcl}, we discuss the advantages of our proposed technique and provide insight into the relationships among different MTM schemes in literature. 

%A well known method for Bayesian estimation is the Metropolis Hastings (MH) algorithm proposed by Metropolis and modified by Hastings ¡¡CITA HASTINGS!!. This algorithm generate a Markov chain whose stationary distribution $p_{e}(x)$ is equal to the target distribution $p_{o}(x)$.

 % The MH algorithm was extended by Liu et al. (2000) ¡¡CITA LIU!!to the case of Multiple-Try Metropolis (MTM). The MTM consists of proposing a fixed number of moves and then selecting one of them with a certain probability (weights). In literature, we can find two interestings papers, in ¡¡CITA ITALIANOS!! we can see a fantastic generalization in which the weight function is not defined. In ¡¡CITA CORRELATED!! presented a extension of MTM in which each point depends of the previous samples. The idea is generated better points in the same iteration, i.e, the  news parallel chains depends the previous chains, and the reference points depends the previous points. This paper present a simple idea, mix the correlation idea with a generic weights.
%%%%%%%%%%%%%%

\section{Multi-point Metropolis algorithm}
\label{SectMPM}
In the classical MH algorithm, a new possible state is drawn from the proposal pdf and the movement is accepted with a suitable decision rule. In the multi-point approach, several correlated samples are generated and, from these, a ``good'' one is chosen.

Specifically, consider a target pdf $p_o(x)$ known up to a constant (hence, we can evaluate $p(x)\propto p_o(x)$). Given a current state $x\in \mathbb{R}$ (we assume scalar values only for simplicity in the treatment), we draw $N$ correlated samples each step from a sequence of different proposal pdfs $\{\pi_j\}_{j=1}^N$, i.e.,
\begin{gather}
\begin{split}
y_1\sim \pi_1(\cdot|x), &y_2\sim \pi_2(\cdot|x,y_1), y_3\sim \pi_3(\cdot|x,y_1,y_2), \ldots \\
& \ldots y_N\sim \pi_N(\cdot|x,y_1,...,y_{N-1}).
\end{split}
\end{gather}
Therefore, we can write the joint distribution of the generated samples as
\begin{equation}
\label{EqQ1}
\small
q_N(y_1,...,y_N|x)=q_N(y_{1:N}|x)=\pi_1(y_1|x)\pi_2(y_2|x,y_1)\cdots \pi_N(y_N|x,y_{1:N-1}),
\end{equation}
i.e., %\marginpar{mas conciso asi, no?}
\begin{equation}
\label{EqQ2}
	q_N(y_{1:N}|x) = \pi_1(y_1|x) \prod_{j=2}^N \pi_j(y_j|x,y_{1:j-1})
\end{equation}
where, for brevity, we use the notation $y_{1:j}\dfn [y_1,...,y_j]$ %for a generic value $j\in \mathbb{N}^+$, 
and $y_{j:1}\dfn[y_j,...,y_1]$ denotes the vector with the reverse order.

A ``good'' candidate among the generated samples is chosen according to %\marginpar{por que no usar $z_{1:m+1}$ si lo acabas de definir?}
%Therefore, 
%let $p(x)$ a function proportional to our target pdf $p_o(x)$ (i.e., $p(x)\propto p_o(x)$), 
weight functions 
$$\omega_j(z_1,...,z_{j+1})\in \mathbb{R}^{j+1}\rightarrow \mathbb{R}^{+}$$ 
where $z_1,...,z_{j+1}$,  are generic variables and $j=1,...,N$. The specific analytic form of the weights needed in this technique is 
\begin{gather}
\label{DefW}
\begin{split}
%&\omega_m(z_1,...,z_{m+1})=p(z_1)q_m(z_{2:m+1}|x) \lambda(z_1,....,z_{m+1}),\\
&\omega_j(z_1,...,z_{j+1})\dfn p(z_1)\pi_1(z_2|z_1)\cdots \pi_N(z_j|z_{1:j-1}) \lambda_j(z_1,....,z_{j+1}),
\end{split}
\end{gather}
where $p(x)\propto p_o(x)$ is the target pdf, $\lambda_j$ can be any bounded, positive, and {\it sequentially symmetric} function, i.e., %\marginpar{y aqui $z_{1:m+1}$ y $z_{m+1:1}$}
\begin{equation}
\lambda_j(z_1,z_{2:j+1})=\lambda_j(z_{j+1:2},z_1).
\end{equation}
Note that, since $q_j(z_{2:j+1}|z_1)=\pi_1(z_2|z_1)\cdots \pi_N(z_j|z_{1:j-1})$ (see Eq.~(\ref{EqQ1})), we can rewrite the weight functions as
\begin{gather}
\label{DefWcorta}
\begin{split}
%&\omega_m(z_1,...,z_{m+1})=p(z_1)q_m(z_{2:m+1}|x) \lambda(z_1,....,z_{m+1}),\\
\omega_j(z_1,...,z_{j+1})= p(z_1)q_j(z_{2:j+1}|z_1) \lambda_j(z_1,z_{2:j+1}).
\end{split}
\end{gather}

%Given a function proportional to our target pdf $p(x)\propto p_o(x)$,we consider the functions $\omega(z_{1:m+1})\in \mathbb{R}^{m+1}\rightarrow \mathbb{R}^{+}$
%\begin{equation}
%\label{DefW}
%\omega_m(z_1,...,z_{m+1})=p(z_1)\pi_1(z_2|z_1)\pi_2(z_3|z_1,z_2)\cdots \pi_N(z_m|z_{1:m-1}).
%\end{equation}
\subsection{Algorithm}
\label{alg:algorithm}
Given a current state $x=x_{t}$, the multi-point Metropolis algorithm consists of the following steps:  
\begin{enumerate}
\item Draw $N$ samples $y_{1:N}=[y_1,y_2,...,y_N]$ from the joint pdf
\begin{equation}
\nonumber
q_N(y_{1:N}|x) = \pi_1(y_1|x) \prod_{j=2}^N \pi_j(y_j|x,y_{1:j-1})
%q(y_{1:N}|x)=\pi_1(y_1|x)\pi_2(y_2|x,y_1)\pi_2(y_3|x,y_1,y_2)\cdots \pi_N(y_N|x,y_{1:N-1}),
\end{equation}
namely, draw $y_j$ from $\pi_j(\cdot|x,y_{1:j-1})$, with $j=1,...,N$. 
\item Calculate the weights $\omega_j(y_{j:1},x)$ as in Eq.~(\ref{DefW}), and normalize them to obtain $\bar{\omega}_j$, $j=1,...,N$. %\marginpar{es necesario que $y_{j:1}$ es al reves?} 
\item Draw a $y=y_k\in\{y_1,....,y_N\}$ according to their weights $\bar{\omega}_1,...,\bar{\omega}_N$.  % \marginpar{3. A usar $\bar{\omega}_{1:N}$ ?} 
\item Set 
\begin{equation}
x_{1}^{*}=y_{k-1}, x_{2}^{*}=y_{k-2}, \ldots, x_{k-1}^{*}=y_{1},
\end{equation}
%\begin{gather}
%\begin{split}
%&x_{1}^{*}=y_{k-1}, \mbox{  }\mbox{  }\mbox{  }\mbox{  }\mbox{  } \mbox{  }\mbox{  }\mbox{  }\mbox{  }\mbox{  } \mbox{  }\mbox{  }\mbox{  }\mbox{  }\mbox{  } \mbox{  }\mbox{  }\mbox{  }\mbox{  }\mbox{  }\\
%&x_{2}^{*}=y_{k-2}, \mbox{  }\mbox{  }\mbox{  }\mbox{  }\mbox{  }\mbox{  }\mbox{  }\mbox{  }\mbox{  }\mbox{  } \mbox{  }\mbox{  }\mbox{  }\mbox{  }\mbox{  }\mbox{  }\mbox{  }\mbox{  }\mbox{  }\mbox{  }\\
%& \vdots \mbox{  }\mbox{  }\mbox{  }\mbox{  }\mbox{  }\mbox{  }\mbox{  }\mbox{  }\mbox{  }\mbox{  } \mbox{  }\mbox{  }\mbox{  }\mbox{  }\mbox{  }\mbox{  }\mbox{  }\mbox{  }\mbox{  }\mbox{  }\\
%&x_{k-1}^{*}=y_{1}, \mbox{  }\mbox{  }\mbox{  }\mbox{  }\mbox{  }\mbox{  }\mbox{  }\mbox{  }\mbox{  }\mbox{  } \mbox{  }\mbox{  }\mbox{  }\mbox{  }\mbox{  }\mbox{  }\mbox{  }\mbox{  }\mbox{  }\mbox{  }\\
%\end{split}
%\end{gather}
%$$x_{1}^{*}=y_{k-1}, x_{2}^{*}=y_{k-2},....,x_{k-1}^{*}=y_{1},$$ 
and finally $x_{k}^{*}=x$. Then, draw other ``reference'' samples 
\begin{equation}
x_{j}^{*}\sim \pi_i(\cdot|y,x^{*}_{1:j-1}),
\end{equation}
for $j=k+1,...,N$. Note that for $j=k+1$ we have 
$$ \pi_j(\cdot|y,x^{*}_{1:j-1})=\pi_j(\cdot|y,x^{*}_{1}=y_{k-1},...,x^{*}_{k-1}=y_{1},x^{*}_{k}=x),$$
and, for $j=k+2,...,N$, we have 
$$\pi_j(\cdot|y,x^{*}_{1:j-1})=\pi_j(\cdot|y,x^{*}_{1}=y_{k-1},...,x^{*}_{k-1}=y_{1},x^{*}_{k}=x, x^{*}_{k+1:j-1}).$$

%for $j=k+1,...,N$. {\color{blue} Note that for $j=k+1$ we have 
%$$ \pi_j(\cdot|y,x^{*}_{1:j-1})=\pi_j(\cdot|y,x^{*}_{1},...,x^{*}_{k-1},x^{*}_{k}=x),$$
%where $x_{k}^{*}$ as in Eq~(7) and, for $j=k+2,...,N$, we have 
%$$\pi_j(\cdot|y,x^{*}_{1:j-1})=\pi_j(\cdot|y,x^{*}_{1}=y_{k-1},...,x^{*}_{k-1}=y_{1},x^{*}_{k}=x, x^{*}_{k+1:j-1}).$$}

\item Compute $\omega_j(x_{j:1}^{*},y)$ as in Eq.~(\ref{DefW}).
\item Let $x_{t+1}=y_k$ with probability
\begin{equation}
\label{alpha1}
\alpha=\min\left[1,\frac{\sum_{j=1}^{N} \omega_j(y_{j:1},x)}{\sum_{j=1}^{N} \omega_j(x_{j:1}^{*},y)}\right],
\end{equation}
otherwise set $x_{t+1}=x$ with probability $1-\alpha$.
\item Set $t=t+1$ and repeat from step 1.
\end{enumerate}
The kernel of this technique satisfies the detailed balance condition as shown in \cite{Qin01}. However, to fulfill this condition, the algorithm needs that the weights are defined \emph{exactly} with the form in Eq.~(\ref{DefW}).

\section{Extension with generic weight functions}
\label{OurSect}

Now, we consider generic weight functions
%As already mentioned, we need to define a generic function of weights that the only restriction is that this function is 
$\omega_j(z_1,...,z_{j+1})\in \mathbb{R}^{j+1}\rightarrow \mathbb{R}^{+},$
that have to be (a) bounded and (b) positive.
%\begin{itemize}
%\item[1)] bounded and
%\item[2)] positive.
%\end{itemize}
In this case, the algorithm can be described as follows. %The first two steps are identical to the algorithm in Subsection \ref{alg:algorithm}. Then, 
\begin{enumerate}
\item Draw $N$ samples $y_{1:N}=[y_1,y_2,...,y_N]$ from the joint pdf
\begin{equation}
\nonumber
q_N(y_{1:N}|x) = \pi_1(y_1|x) \prod_{j=2}^N \pi_j(y_j|x,y_{1:j-1})
%q(y_{1:N}|x)=\pi_1(y_1|x)\pi_2(y_2|x,y_1)\pi_2(y_3|x,y_1,y_2)\cdots \pi_N(y_N|x,y_{1:N-1}),
\end{equation}
namely, draw $y_j$ from $\pi_j(\cdot|x,y_{1:j-1})$, with $j=1,...,N$. 
\item Choose some suitable (bounded and positive) weight functions. Then, calculate each weight $\omega_j(y_{j:1},x)$, and normalize them to obtain $\bar{\omega}_j$, $j=1,...,N$. %\marginpar{es necesario que $y_{j:1}$ es al reves?} 
	\item Draw a $y=y_k\in\{y_1,....,y_N\}$ according to  $\bar{\omega}_1,...,\bar{\omega}_N$, and set $\bar{W}_y=\bar{\omega}_k$, i.e.,
	\begin{equation}
	\label{OmegaYluca}
	\bar{W}_y\dfn \frac{\omega_k(y_{k:1},x)}{\sum_{j=1}^{N}\omega_j(y_{j:1},x)}.
	\end{equation}
	\item Set  
	\begin{equation}
	   \label{EqJesselikes}
           x_{1}^{*}=y_{k-1}, x_{2}^{*}=y_{k-2}, \ldots, x_{k-1}^{*}=y_{1},
          \end{equation}
%	\begin{gather}
%	\label{EqJesselikes}
%            \begin{split}
%            &x_{1}^{*}=y_{k-1}, \mbox{  }\mbox{  }\mbox{  }\mbox{  }\mbox{  } \mbox{  }\mbox{  }\mbox{  }\mbox{  }\mbox{  } \mbox{  }\mbox{  }\mbox{  }\mbox{  }\mbox{  } \mbox{  }\mbox{  }\mbox{  }\mbox{  }\mbox{  }\\
%           &x_{2}^{*}=y_{k-2}, \mbox{  }\mbox{  }\mbox{  }\mbox{  }\mbox{  }\mbox{  }\mbox{  }\mbox{  }\mbox{  }\mbox{  } \mbox{  }\mbox{  }\mbox{  }\mbox{  }\mbox{  }\mbox{  }\mbox{  }\mbox{  }\mbox{  }\mbox{  }\\
%            & \vdots \mbox{  }\mbox{  }\mbox{  }\mbox{  }\mbox{  }\mbox{  }\mbox{  }\mbox{  }\mbox{  }\mbox{  } \mbox{  }\mbox{  }\mbox{  }\mbox{  }\mbox{  }\mbox{  }\mbox{  }\mbox{  }\mbox{  }\mbox{  }\\
%           &x_{k-1}^{*}=y_{1}, \mbox{  }\mbox{  }\mbox{  }\mbox{  }\mbox{  }\mbox{  }\mbox{  }\mbox{  }\mbox{  }\mbox{  } \mbox{  }\mbox{  }\mbox{  }\mbox{  }\mbox{  }\mbox{  }\mbox{  }\mbox{  }\mbox{  }\mbox{  }\\
%         \end{split}
%          \end{gather}
          and finally $x_{k}^{*}=x$. Then, draw the remaining ``reference'' samples 
           \begin{equation}
                x_{j}^{*}\sim \pi_j(\cdot|y,x^{*}_{1:j-1}),
               \end{equation}
                  for $j=k+1,...,N$. 
	%$$x_{1}^{*}=y_{k-1}, x_{2}^{*}=y_{k-2},....,x_{k-1}^{*}=y_{1} \mbox{  } \mbox{ and }\mbox{  } x_{k}^{*}=x.$$ 	
	\item Compute the general weights $\omega_j(x_{j:1}^{*},y)$ and calculate the normalized weight
	\begin{equation}
	\label{OmegaXluca}
	\bar{W}_x\dfn \frac{\omega_k(x^{*}_{k:1},y)}{\sum_{j=1}^{N}\omega_j(x^{*}_{j:1},y)}.
	\end{equation}
	\item Set $x_{t+1}=y_k$ with probability
	%%%% CAMBIO AQUI ADD A FORMULA %%%%
	\begin{equation}
	\label{sectalpha2}
	\small
	\alpha=\min\left[1,\frac{p(y)\pi_1(x^{*}_1|y)\pi_2(x^{*}_2|y,x^{*}_1) \cdots \pi_k(x^{*}_k|y,x^{*}_1,...,x^{*}_{k-1})}{p(x)\pi_1(y_1|x)\pi_2(y_2|x,y_1) \cdots \pi_k(y_k|x,y_1,...,y_{k-1})}\frac{\bar{W}_x}{\bar{W}_y}\right].
	\end{equation}
	We can rewrite it in a more compact form as
	\begin{equation}
	\label{sectalpha}
	\alpha=\min\left[1,\frac{p(y)q_k(x^{*}_{1:k}|y)}{p(x)q_k(y_{1:k}|x)}\frac{\bar{W}_x}{\bar{W}_y}\right],
	\end{equation}
	where we recall 
	\begin{equation}
	\label{otravezQ}
	q_k(y_{1:k}|x)=\pi_1(y_1|x) \prod_{j=2}^{k}\pi_j(y_j|x,y_{1:j-1}).
	\end{equation}
	where $k$ is the index of the chosen sample $y_k$.\\
	Otherwise, set $x_{t+1}=x$ with probability $1-\alpha$.
	\item Set $t=t+1$ and repeat from step 1.
	\end{enumerate}
	We emphasise that in the algorithm above we have not specifically defined the weight functions.  
	%%% FRASE AÑADIDA %%%%

%%%%%%%%%%%%%%%%%%%%%%%%%%%%%%
\subsection{Examples of weight functions} 
%%%%%%%%%%%%%%%%%%%%%%%%%%%%%%
The weight functions must to be bounded and positive. The choice can depend on some criteria such as improving performance or reducing computational complexity. 
If the target density is bounded, two possibilities are
\begin{equation}
\omega_j(z_1,...,z_{j+1})=p(z_1), 
\end{equation}
or 
\begin{equation}
\omega_j(z_1,...,z_{j+1})=p(z_1)p(z_2)\cdots p(z_{j+1}),
\end{equation}
with $j=1,...,N$. Another possible choices are the following
\begin{equation}
\omega_j(z_1,...,z_{j+1})=\left[\frac{p(z_1)}{q_j(z_{1:j}|z_{j+1})}\right]^{\theta},
\end{equation}
where $\theta>0$ is a positive constant, or
\begin{equation}
\omega_j(z_1,...,z_{j+1})=
\frac{p(z_j)}{q_1(z_{j}|z_{j+1})} \frac{p(z_{j-1})}{q_2(z_{j-1:j}|z_{j+1})} \cdots \frac{p(z_1)}{q_j(z_{1:j}|z_{j+1})},
\end{equation}
and a third possible choice
\begin{equation}
\omega_j(z_1,...,z_{j+1})=\frac{p(z_1)}{\pi_{j+1}(z_1|z_{j+1:2})},
\end{equation}
where $\pi_{j+1}(z_1|z_{j+1:2})$ is the $j+1$-th proposal pdf used in the step 1 of the algorithm.
It is important to remark that the $z$-variables are ordered such that $z_1$ is the most recently generated sample, $z_j$ is the first drawn sample, and $z_{j+1}$ represents the \emph{previous step} of the chain.

Clearly, owing to the great flexibility in the construction of the weight functions, it can be difficult to assert which is the best choice in terms of performances of the algorithm. However, evidently, in general including more statistical information in the weights can improve performance yet, at the same time, increases the computational cost of the designed technique. 

More specific theoretical or empirical studies are needed to clear up this issue. Indeed, observe that the point of the best selection of the weights is even unclear in the classical MTM by \cite{Liu00}, as for the method in \cite{Pandolfi10}, for instance.

%Usually the importance weights provide better performances....

%Some possibilities are shown in Table \ref{}.

%%%%%%%%%%%%%%%%%%%%%%%%%%%%%
\subsection{Relationship with the independent multiple tries scheme} 
In \cite{Pandolfi10} i.i.d.~candidates are proposed at each time step.  
The acceptance probability $\alpha$ in Eqs.~(\ref{sectalpha2})-(\ref{sectalpha}) may appear similar to the acceptance probability in \cite{Pandolfi10}.  
However, note that the expression of $\alpha$ in Eq.~(\ref{sectalpha}) is different to the acceptance probability in \cite{Pandolfi10} for two main reasons: 
\begin{itemize}
\item[(a)] the first factor $\frac{p(y)q_k(x^{*}_{1:k}|y)}{p(x)q_k(y_{1:k}|x)}$ is distinct (see Eqs.\ (\ref{otravezQ})), and  
\item[(b)] the definition and computation of $\bar{W}_x$ and $\bar{W}_y$ (see Eqs.\ (\ref{OmegaYluca}) and (\ref{OmegaXluca})) are also different since here the weight functions take in account the previous generated samples (in the same time step). 
\end{itemize}

If here we set $\pi_j(y_j|x,y_{1:j-1})= \pi(y_j|x)$  for all $j=1,...,N$, then the steps of our algorithm coincides exactly with those of technique in \cite{Pandolfi10} {\em except for} the step 4. Indeed, the way of choosing the ``reference'' points are different in the two methods (in our case, some of them are fixed while in \cite{Pandolfi10} all the reference point are chosen random). We can find a specular difference between the methods in \cite{Liu00} and \cite{Qin01}. 

%%%%%%%%%%%%%%%%%%%%%%%%%%%%%%%%%%%%%%%%

 %%%%%%%%%%%%%%%%%%%%%%%%%%
\subsection{Multi-point Metropolis as specific case}
%%%%%%%%%%%%%%%%%%%%%%%%%%
\label{SpecCase}
In the case when the weight functions are chosen as in Eq.~(\ref{DefWcorta}), i.e., $\omega_k(z_1,...,z_{j+1})=p(z_1)q_j(z_{2:j+1}|z_1) \lambda_j(z_1,....,z_{j+1}),$
%\begin{gather}
%\begin{split}
%\label{W2}
%&\omega_m(z_1,...,z_{m+1})=p(z_1)q_m(z_{2:m+1}|x) \lambda(z_1,....,z_{m+1}),\\
%\omega_k(z_1,...,z_{j+1})=p(z_1)q_j(z_{2:j+1}|z_1) \lambda_j(z_1,....,z_{j+1}),
%\end{split}
%\end{gather}
where 
\begin{equation}
\label{SeqSymm}
\lambda_j(z_1,z_{2:j+1})=\lambda_j(z_{j+1:2},z_1),
\end{equation}
is sequentially symmetric, then our scheme coincides exactly with the standard multi-point Metropolis method in \cite{Qin01}.
Indeed, first of all we can rewrite the expression (\ref{sectalpha}) as 
\begin{equation}
\label{alpha3}
\alpha=\min\left[1,\frac{p(y)q_k(x^{*}_{1:k}|y)}{p(x)q_k(y_{1:k}|x)}\frac{\omega_k(x^{*}_{k:1},y)}{\omega_k(y_{k:1},x)} \frac{\sum_{j=1}^{N} \omega_j(y_{j:1},x)}{\sum_{j=1}^{N} \omega_j(x_{j:1}^{*},y)}\right].
\end{equation}
Then, recalling the Eq.~(\ref{EqJesselikes}), i.e., $x_{1}^{*}=y_{k-1}, x_{2}^{*}=y_{k-2},....,x_{k-1}^{*}=y_{1},$
 $x_{k}^{*}=x$ and $y=y_k$,  the two  weights $\omega_k(x^{*}_{k:1},y)$ and $\omega_k(y_{k:1},x)$  can be expressed exactly as 
 \begin{gather}
 \nonumber
 \begin{split}
 \omega_k(x^{*}_{k:1},y)&=\omega_k(x^{*}_{k}=x,x^{*}_{k-1}=y_1,...,x^{*}_{1}=y_{k-1},y=y_k)\\
 &=p(x)q_k(y_{1:k}|x) \lambda_k(x,y_{1:k}),\\
\end{split}
\end{gather}
and
\begin{gather}
 \nonumber
 \begin{split}
\omega_k(y_{k:1},x)&=\omega_k(y_{k}=y,y_{k-1}=x^{*}_{1},...,y_{1}=x^{*}_{k-1},x=x^{*}_k)\\
&=p(y)q_k(x^{*}_{1:k}|y)\lambda_k(y,x^{*}_{1:k}),
\end{split}
\end{gather}
respectively.  Therefore
 replacing the weights $\omega_k(x^{*}_{k:1},y)$ and $\omega_k(y_{k:1},x)$ in Eq.~(\ref{alpha3}), we obtain 
\begin{equation}
\nonumber
\alpha=\min\left[1,\frac{\lambda_k(x,y_{1:k})}{\lambda_k(y,x^{*}_{1:k})}\frac{\sum_{j=1}^{N} \omega_j(y_{j:1},x)}{\sum_{j=1}^{N} \omega_j(x_{j:1}^{*},y)}\right]=\min\left[1,\frac{\sum_{j=1}^{N} \omega_j(y_{j:1},x)}{\sum_{j=1}^{N} \omega_j(x_{j:1}^{*},y)}\right],
\end{equation}
that coincides with acceptance probability in Eq.~(\ref{alpha1}) of the standard multi-point Metropolis algorithm. Note that we have considered $\lambda_k(x,y_{1:k})=\lambda_k(y,x^{*}_{1:k})$. Indeed, since $x^*_k=x$ we can write    
$\lambda_k(x,y_{1:k})=\lambda_k(y,x^{*}_{1:k-1},x),$
then because $x^{*}_{1:k-1}=y_{k-1:1}$, we obtain
$\lambda_k(x,y_{1:k})=\lambda_k(y,y_{k-1:1},x),$
and as $y=y_k$, finally  we have
$$\lambda_k(x,y_{1:k})=\lambda_k(y_{k:1},x),$$
that is exactly the condition assumed in Eq. (\ref{SeqSymm}).
%%%%%%%%%%%%%%%%%%%%%%%%%%%%
%\subsection{Observation about the reference point ??} 
 %%%%%%%%%%%%%%%%%%%%%%%%%%%%
%\marginpar{a notar, que si usamos esta seccion, tenemos que modificar un par de referencias de la Introduction}
 %%%%%%%%%%%%%%%%%%%%%%%%%%%%
%%%%%%%%%%%%%%%%%    
In the following, we show the proposed technique satisfies the detailed balance condition.    

%%%%% comentar que anche los pesos dependen de las muestras anteriores %%%

\section{Proof of the detailed balance condition}
\label{SectProof}
To guarantee that a Markov chain generated by an MCMC method converges to the target distribution $p(x)\propto p_o(x)$, the kernel $A(y|x)$ of the corresponding algorithm fulfills the following detailed balance condition\footnote{Note that the detailed balance condition is sufficient but not necessary condition. Namely, the detailed balance ensures invariance. The converse is not true. Markov chains that satisfy the detailed balance condition are called {\it reversible}.}   
$$p(x)A(y|x)=p(y)A(x|y).$$
First of all, we have to find the kernel $A(y|x)$ of the algorithm, i.e., the conditional probability to move from $x$ to $y$. For simplicity, we consider the case $x\neq y$ (case $x=y$ is trivial). The kernel can be expressed as
 \begin{equation}
A(y=y_k|x)=\sum_{j=1}^{N} h(y=y_k|x,k=j),
\end{equation}
where $h(y=y_k|x,k=j)$ is the probability of accepting $x_{t+1}=y_k$ given $x_{t}=x$ when the chosen sample $y_k$ is the $j$-th candidate, i.e., when $y_k=y_j$.

In the sequel, we study just one $h(y=y_k|x,k)$ for a generic $k\in\{1,...,N\}$. Indeed,  if $h(y=y_k|x,k)$ fulfills the detailed balance condition (it is symmetric w.r.t.\ $x$ and $y$), then  $A(y|x)$ also satisfies the detailed balance because it is a sum of symmetric functions. Therefore, we want to show that 
$$p(x)h(y|x,k)=p(y)h(x|y,k),$$
for a generic $k\in\{1,...,N\}$. Following the steps above of the algorithm, we can write 
\begin{gather}
%\small
\nonumber
\begin{split}
 &p(x)h(y|x,k)= \\
&=p(x) \int\cdots\int \left[\prod_{j=1}^{N}\pi_j(y_j|x,y_{1:j-1})\right] \frac{\omega_k(y_{k:1},x)}{\sum_{j=1}^{N}\omega_j(y_{j:1},x)}\left[\prod_{i=k+1}^{N}\pi_{i}(x^{*}_i|y,x^{*}_{1:i-1})\right]\cdot  \\ 
 &\mbox{ }\mbox{ }\mbox{ }\mbox{ }\mbox{ }\mbox{ }\mbox{ }\mbox{ }\mbox{ }\mbox{ }\mbox{ }\mbox{ }\mbox{ }\mbox{ }\mbox{ }\mbox{ }\mbox{ }\mbox{ }\mbox{ }\mbox{ }\cdot  \min\left[1,\frac{p(y)q_k(x^{*}_{1:k}|y)}{p(x)q_k(y_{1:k}|x)}\frac{\bar{W}_x}{\bar{W}_y}\right] dy_{1:k-1}dy_{k+1:N}dx^{*}_{k+1:N}.
\end{split}
\end{gather}
Note that each factor inside the integral corresponds to a step of the method described in the previous section. The integral is over all auxiliary variables. 
Recalling the definition of the joint probability $q_k(y_{1:k}|x)$ and $\bar{W}_y$, the expression can be simplified to
\begin{gather}
\small
\nonumber
\begin{split}
 &p(x)h(y|x,k)= \\
&=p(x) \int\cdots\int q_k(y_{1:k}|x)\cdot \left[\prod_{j=k+1}^{N}\pi_j(y_j|x,y_{1:j-1})\right]\cdot \bar{W}_y\cdot \left[\prod_{i=k+1}^{N}\pi_{i}(x^{*}_i|y,x^{*}_{1:i-1})\right]\cdot  \\ 
 &\mbox{ }\mbox{ }\mbox{ }\mbox{ }\mbox{ }\mbox{ }\mbox{ }\mbox{ }\mbox{ }\mbox{ }\mbox{ }\mbox{ }\mbox{ }\mbox{ }\mbox{ }\mbox{ }\mbox{ }\mbox{ }\mbox{ }\mbox{ }\min\left[1,\frac{p(y)q_k(x^{*}_{1:k}|y)}{p(x)q_k(y_{1:k}|x)}\frac{\bar{W}_x}{\bar{W}_y}\right] dy_{1:k-1}dy_{k+1:N}dx^{*}_{k+1:N}, 
\end{split}
\end{gather}
and we only arrange it, obtaining 
\begin{gather}
\small
\nonumber
\begin{split}
 &p(x)h(y|x,k)= \\
&= \int\cdots\int p(x)q_k(y_{1:k}|x)\bar{W}_y\left[\prod_{j=k+1}^{N}\pi_j(y_j|x,y_{1:j-1})\right] \left[\prod_{i=k+1}^{N}\pi_{i}(x^{*}_i|y,x^{*}_{1:i-1})\right]\cdot  \\ 
 &\mbox{ }\mbox{ }\mbox{ }\mbox{ }\mbox{ }\mbox{ }\mbox{ }\mbox{ }\mbox{ }\mbox{ }\mbox{ }\mbox{ }\mbox{ }\mbox{ }\mbox{ }\mbox{ }\mbox{ }\mbox{ }\mbox{ }\mbox{ }\min\left[1,\frac{p(y)q_k(x^{*}_{1:k}|y)}{p(x)q_k(y_{1:k}|x)}\frac{\bar{W}_x}{\bar{W}_y}\right] dy_{1:k-1}dy_{k+1:N}dx^{*}_{k+1:N}.
\end{split}
\end{gather}
Now, we multiply the two members of the function $\min[\cdot,\cdot]$ by the factor $p(x)q_k(y_{1:k}|x)\bar{W}_y$ so that    
\begin{gather}
\nonumber
\begin{split}
 &p(x)h(y|x,k)=  \int\cdots\int \left[\prod_{j=k+1}^{N}\pi_j(y_j|x,y_{1:j-1})\right] \left[\prod_{i=k+1}^{N}\pi_{i}(x^{*}_i|y,x^{*}_{1:i-1})\right]\cdot  \\ 
 &\mbox{ }\mbox{ }\mbox{ }\mbox{ }\mbox{ }\mbox{ }\mbox{ }\mbox{ }\mbox{ }\mbox{ }\mbox{ }\mbox{ }\mbox{ }\mbox{ }\mbox{ }\mbox{ }\mbox{ }\mbox{ }\mbox{ }\mbox{ }\min\left[p(x)q_k(y_{1:k}|x)\bar{W}_y,p(y)q_k(x^{*}_{1:k}|y)\bar{W}_x\right] dy_{1:k-1}dy_{k+1:N}dx^{*}_{k+1:N}.
\end{split}
\end{gather}
Therefore, it is straightforward that the expression above is symmetric in $x$ and $y$. Indeed, we can exchange the notations of $x$ and $y$, and  $x^{*}_i$ and $y_j$, respectively, and the expression does not vary. Then we can write
\begin{equation}
p(x)h(y|x,k)=p(y)h(x|y,k).
\end{equation}
We can repeat the same development for each $k$ obtaining  
\begin{equation}
p(x)A(y|x)=p(y)A(x|y),
\end{equation}   
that is the detailed balance condition. Therefore, the generated Markov chain converges to our target pdf.

%%%%%%%%%%%%%%%%%%%%
 \section{Toy example} 
 %%%%%%%%%%%%%%%%%%%%
 \label{ToyEx}
 
Now we provide a simple numerical simulation to show an example of multi-point scheme with generic weight functions and compare it with the technique in \cite{Pandolfi10}.   
% We consider a simple example to provide a ``numerical proof''. We recall that the theoretical proof is given in Section \ref{SectProof}.
%Note that in this
Let $X\in \mathbb{R}$ be a random variable\footnote{Note that  we consider a scalar variable only to simplify the treatment. Clearly, all the considerations and algorithms are valid for multi-dimensional variables.}  with bimodal pdf 
\begin{equation}
p_o(x)\propto p(x)=\exp\left\{-(x^2-4)^2/4 \right\}.
\end{equation}
Our goal is to draw samples from $p_o(x)$ using our proposed multi-point technique.  

We consider a Gaussian densities as proposal pdfs (a standard choice)  
\begin{equation}
\pi_j(y_j|x_t,y_{1:j-1}) \propto \exp\left\{-\frac{(y_j-\mu_j)^2}{2\sigma^2}\right\}
\end{equation}
where we use $\sigma^2=1$ and
\begin{equation}
\mu_j=\frac{\gamma_1}{i-1} (x_t+y_1+...+y_{i-2})+\gamma_2 y_{i-1}, 
\end{equation}
i.e, $\mu$ is a weighted mean ($\gamma_1+\gamma_2=1$) of the previous state $x_t$ and the previous generated samples (at the same time step). Specifically, we set $\gamma_1=0.2$ and $\gamma_2=0.8$. 
   
Moreover, we choose very simple weight functions depending only on first variable and on the target pdf
\begin{equation}
\label{FirstW}
\omega_j^{(1)}(z_1,z_2,....,z_{j+1})=[p(z_1)]^{\theta},
\end{equation}
with $\theta=1/2$. Note that $p(\cdot)$ is bounded and also positive (since it is a pdf). This kind of weights cannot be used in the multi-point scheme of \cite{Qin01}, expect for $\theta=1$ and using a specific sequence of the proposal pdfs. Moreover, for $\theta=1$ this weight function can be also used in a standard MTM of \cite{Liu00} if the chosen proposal density $\pi(y|x)$ is symmetric (i.e, $\pi(y|x)=\pi(x|y)$ and choosing $\lambda(x,y)=\frac{1}{\pi(x|y)}$). 

We also compare the performances of the proposed algorithms with the weights as
\begin{equation}
\label{SecondW}
\omega_j^{(2)}(z_1,...,z_{j+1})=p(z_1)p(z_2)\cdots p(z_{j+1}),
\end{equation}
and
\begin{equation}
\label{TerzoW}
\omega_j^{(3)}(z_1,...,z_{j+1})=\frac{p(z_1)}{\pi_{j+1}(z_1|z_{j+1:2})}.
\end{equation} 
Then, we run the proposed multi-point algorithm with different numbers $N$ of candidates and calculate the estimated acceptance rate (the averaged probability of accepting a movement) and linear correlation coefficient (between one state of the chain and the next). We also run the method in \cite{Pandolfi10} with proposal pdf 
$\pi(y_j|x_t)\propto \exp\left\{-\frac{(y_j-x_t)^2}{2\sigma^2}\right\}$ 
and compare the performances, using weight functions as in Eq.~(\ref{FirstW}) and third type in Eq. (\ref{TerzoW}). Because the samples are generated independently, we do not compare using weights in Eq.~(\ref{SecondW}), as statistically this no longer makes sense.

Moreover, observe that in the scheme of \cite{Pandolfi10} (where the candidates are drawn independently), the weight functions in Eq.~(\ref{TerzoW})  become $\omega^{(3)}(y_j,x_{t-1})=\frac{p(y_j)}{\pi(y_j|x_{t-1})}$ where $x_{t-1}$ is the previous step of the chain\footnote{Note that, in the expression of the weights $\omega^{(3)}(y_j,x_{t-1})$, we remove the subscript $j$ because in \cite{Pandolfi10}  the analytic form of the weights is the same for each generated sample $y_j$, $j=1,...,N$.}. Note also that this particular choice of weights $\omega^{(3)}$ can be used in the standard MTM of \cite{Liu00} (by choosing $\lambda(x,y)=\frac{1}{\pi(y|x)\pi(x|y)}$) and, in this case, the technique of \cite{Pandolfi10} coincides with a standard MTM.
	%\footnote{{\color{blue}This particular choice of the weights ($\omega^{(3)}$) is also possible in the standard MTM when the proposal pdf is symmetric. As in this case we are using a symmetric Gaussian proposal , the technique of \cite{Pandolfi10} coincides with a standard MTM.}}

\begin{figure}[htb]
\centerline{
\subfigure[]{\includegraphics[width=4.5cm]{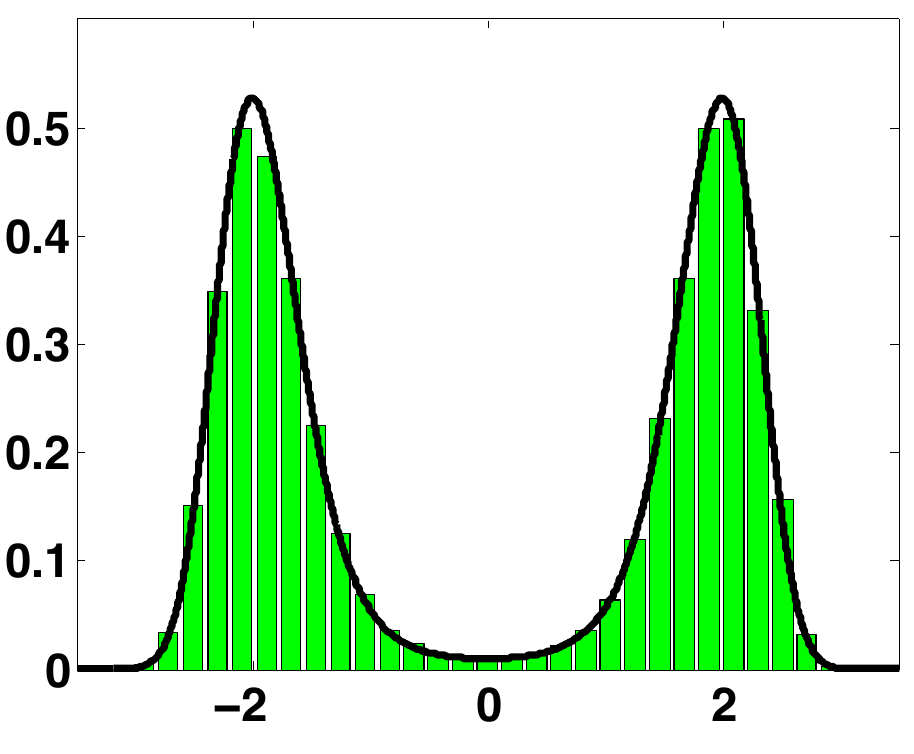}}
%\subfigure[]{\includegraphics[width=4.5cm]{Fig2creo.pdf}}
\subfigure[]{\includegraphics[width=4.5cm]{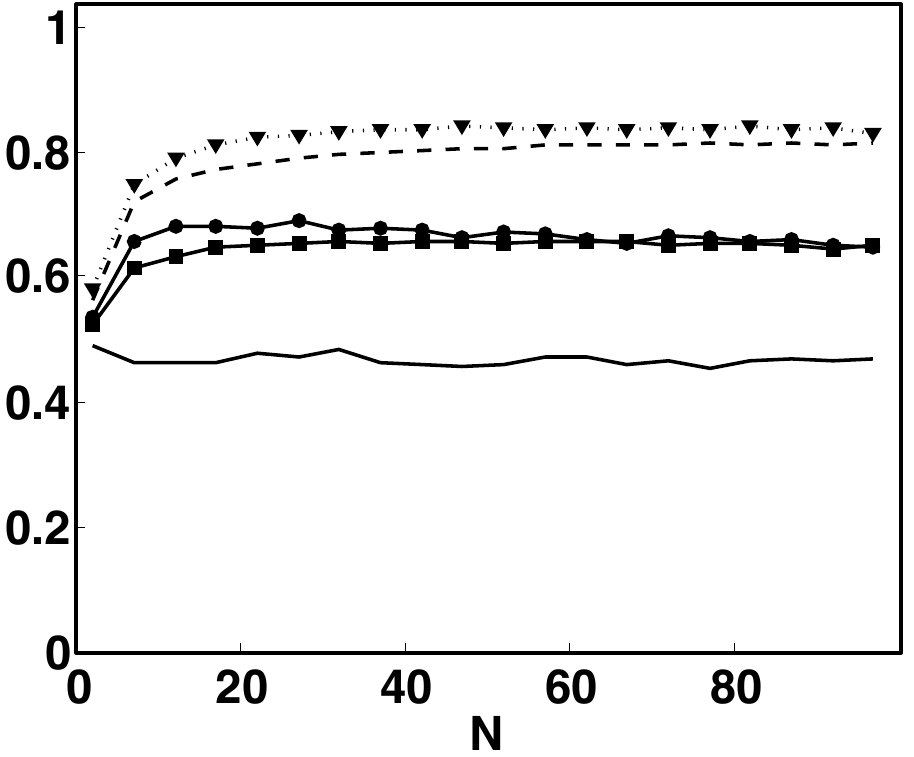}}
%\subfigure[]{\includegraphics[width=4.5cm]{Fig3creo.pdf}}
\subfigure[]{\includegraphics[width=4.5cm]{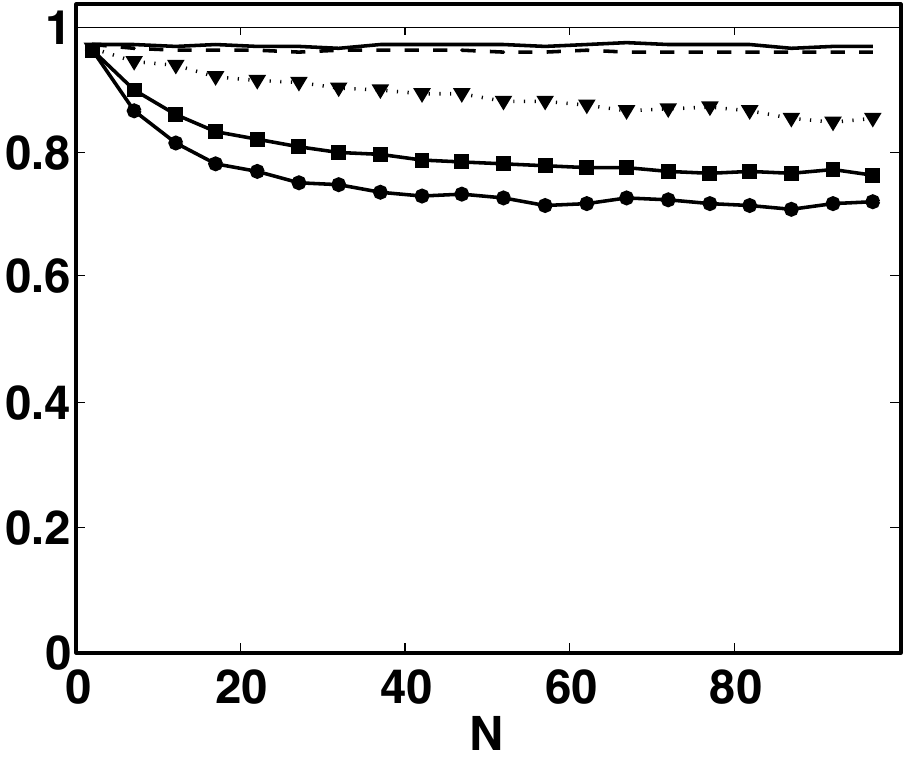}}
 		}
\caption{{\bf (a)} The target density $p_o(x)$ (solid line) and the normalized histogram of the samples generated using the proposed scheme and with $N=10$. {\bf (b)}  The mean acceptance probability of jumping in a new state, depending on the number of tries $N$. We show the results of the technique in \cite{Pandolfi10} (dashed line for weights $\omega_j^{(1)}$ and dotted line with triangles for $\omega_j^{(3)}$) and our method (squares for $\omega_j^{(1)}$, solid line with $\omega_j^{(2)}$, and with circles with $\omega_j^{(3)}$).  {\bf (c)} Estimated linear correlation coefficient  depending on the number of tries $N$ for the different techniques.}
\label{fig1}
\end{figure}

Figure \ref{fig1}(a) depicts the target density $p_o(x)$ (solid line) and the normalized histogram of $100,000$ samples drawn using our proposed scheme and $N=10$. %We can observe that the histogram approximates closely the shape of the target pdf, i.e., the Markov chain generated by the novel scheme converges to $p_o(x)$.
Figures \ref{fig1}(b)-(c) illustrate the mean acceptance probability and the estimated correlation coefficient (for different values of $N$ and averaged using $5,000$ runs) of the two techniques and different choice of weights: our method is shown with squares using $\omega_j^{(1)}$, with solid line using $\omega_j^{(2)}$ and with circles using $\omega_j^{(3)}$. The performances of the method in \cite{Pandolfi10} are depicted with dashed line corresponding to the first choice $\omega_j^{(1)}$, and dotted line with triangles for $\omega_j^{(3)}$. 

We can see although the proposed technique always attains smaller acceptance rates, the resulting correlations are always smaller than the correlations obtained by the other method, except using weights $\omega_j^{(2)}$ in Eq.~(\ref{SecondW}). Moreover, the best results are obtained with the proposed technique using the weights $\omega_j^{(3)}$ in Eq.~(\ref{TerzoW}).
In this case, the correlation decreases when $N$ increases, up to $0.72$ with $N=100$. 

\section{Discussion}
\label{sectConcl}
%The multi-point Metropolis algorithm (\citep{Qin01}) is a generalization of the standard Metropolis-Hasting method that uses several correlated candidates in order to allow large step-size jumps without lowering the acceptance rate. This technique is based on a specific definition of the weight functions.

In this work, we have introduced a Metropolis scheme with multiple correlated points where the weight functions are not defined specifically, i.e., the analytic form can be chosen arbitrarily. 
%The only constraints are that the weights must be bounded and positive functions. This is unlike the existing multi-point Metropolis algorithm which is based on a specific definition of the weight functions. 
We proved that our novel scheme satisfies the detailed balance condition.

Our approach draws from two different approaches \citep{Pandolfi10, Qin01} to form a novel efficient and flexible multi-point scheme.

The multi-point approach with correlated samples provides different advantages over the standard MTM. For instance, the multi-point procedure can iteratively improve the proposal pdfs in two different ways. Firstly, since the proposal pdfs can be distinct, as in \cite{Casarin2011}, it is possible to tune the parameters of each proposal in every time step. % (see for instance \citep{Casarin2011,Andrieu08}). 
 Secondly, since the candidates are generated sequentially, successive proposal pdfs can be improved learning from the previously produced samples during the same time step. 
 
Moreover, in our technique, the only constraints of the weight functions are that they must be bounded and positive, unlike in the existing multi-point Metropolis algorithm \citep{Qin01} which is based on a specific definition of the weight functions. Here the weights can be chosen with respect to some criteria such as improving performance or reducing computational complexity. Thus our method avoids any control or check the existence of a
suitable function $\lambda$ and, therefore, the selection of the weight functions is broader and easier. 

It is interesting to observe that, in general,  the function $\lambda$ may depend on the proposal pdf for a specific choice of weights and, in some cases, may entail certain constraints on the proposal pdf (such as that it be symmetric, for instance). An important consequence of this, it is that the weights can be chosen independently of the specific proposal pdf used in the algorithm. Namely, the proposal distribution and the weight functions can be selected separately, to fit well to the specific problem and to improve the performance of the technique. However, further theoretical or empirical studies are needed to determine the best choice of weight functions given a certain proposal and target density. 

\begin{figure}[h]
\centering{
\includegraphics[width=14cm]{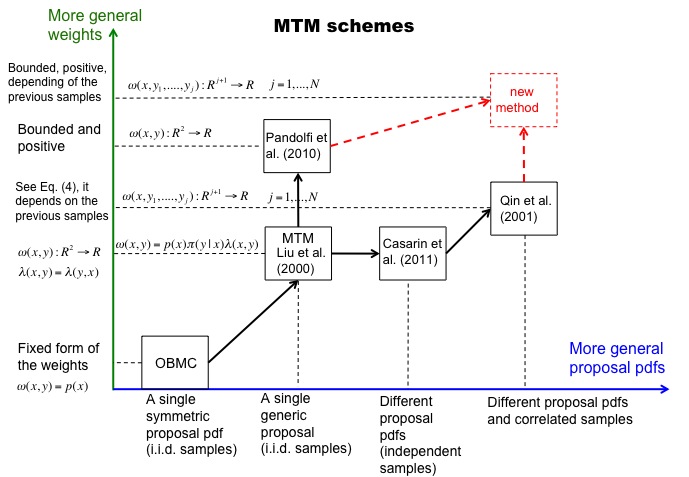}
}
\caption{Comparison of different MTM schemes in literature according to the flexibility in the choice of the proposal and weight functions. With the acronym OBMC we indicate the {\it orientational bias Monte Carlo} introduced by \cite{Frenkel96}.}
\label{FigTeo}
\end{figure}

Furthermore, unlike in \cite{Pandolfi10}, in our method the weights can depend on the previous candidates, and the dimension of the weight functions grows from $\mathbb{R}^2$ to $\mathbb{R}^{N}$, thus being more general and potentially more powerful.
Figure \ref{FigTeo} illustrates the relationships among different MTM schemes according to the flexibility in the choice of the proposal and weight functions. Finally, we have also shown a numerical simulation and a simple multi point scheme that provides good performances reducing the correlation in the produced chain.

\section{Acknowledgements}
{\footnotesize We would like to thank the Reviewer for his comments which have helped us to improve the first version of manuscript.
%This work was partially funded by the Spanish government (Ministerio de Educaci\'{o}n y Ciencia \ldots, \ldots), Universidad Carlos III (\ldots) and the European Union (\ldots).
Moreover, this work has been partially supported by Ministerio de Ciencia e Innovaci\'on of Spain (project MONIN, ref.	TEC-2006-13514-C02- 01/TCM, Program Consolider-Ingenio 2010, ref. CSD2008- 00010 COMONSENS, and Distribuited Learning Communication and Information Processing (DEIPRO) ref. TEC2009-14504-C02-01) and Comunidad Autonoma de Madrid (project PROMULTIDIS-CM, ref. S-0505/TIC/0233). }

\bibliographystyle{elsarticle-harv}
\bibliography{bibliografia} 

%\bibliographystyle{IEEEbib}
%\bibliographystyle{plain}

%% Authors are advised to submit their bibtex database files. They are
%% requested to list a bibtex style file in the manuscript if they do
%% not want to use elsarticle-harv.bst.

%% References without bibTeX database:

% \begin{thebibliography}{00}

%% \bibitem must have one of the following forms:
%%   \bibitem[Jones et al.(1990)]{key}...
%%   \bibitem[Jones et al.(1990)Jones, Baker, and Williams]{key}...
%%   \bibitem[Jones et al., 1990]{key}...
%%   \bibitem[\protect\citepauthoryear{Jones, Baker, and Williams}{Jones
%%       et al.}{1990}]{key}...
%%   \bibitem[\protect\citepauthoryear{Jones et al.}{1990}]{key}...
%%   \bibitem[\protect\astroncite{Jones et al.}{1990}]{key}...
%%   \bibitem[\protect\citepname{Jones et al., }1990]{key}...
%%   \harvarditem[Jones et al.]{Jones, Baker, and Williams}{1990}{key}...
%%

% \bibitem[ ()]{}

% \end{thebibliography}

\end{document}